\documentclass[a4paper,twocolumn]{article}
\usepackage{geometry}
\usepackage[utf8]{inputenc}
\usepackage{authblk}
\usepackage{graphicx}
\graphicspath{{figs/}}
\usepackage{cite}
\usepackage[font=small]{caption}
\usepackage{eurosym}
\usepackage{hyperref}
\usepackage{dblfloatfix} 

\newcommand{\sitablename}{Supplementary Table}
\newcommand{\sifigname}{Supplementary Figure}
\newcommand{\sinotename}{Supplementary Note}

\title{Growing Urban Bicycle Networks}
\author[a,b,c]{Michael Szell\thanks{}}
\author[d]{Sayat Mimar}
\author[d]{Tyler Perlman}
\author[d]{Gourab Ghoshal}
\author[a,b,c,e]{\mbox{Roberta Sinatra}}
\affil[a]{NEtwoRks, Data, and Society (NERDS), IT University of Copenhagen, 2300 Copenhagen, Denmark}
\affil[b]{Complexity Science Hub Vienna, 1080 Vienna, Austria}
\affil[c]{ISI Foundation, 10126 Turin, Italy}
\affil[d]{Department of Physics and Astronomy,
University of Rochester, Rochester, NY 14627, USA}
\affil[e]{Copenhagen Center for Social Data Science (SODAS), University of Copenhagen, 1353 Copenhagen, Denmark}

\date{}

\begin{document}

\twocolumn[
  \begin{@twocolumnfalse}

\maketitle

\vspace{-0.4cm}
\begin{abstract}
\noindent
Cycling is a promising solution to unsustainable car-centric urban transport systems. However, prevailing bicycle network development follows a slow and piecewise process, without taking into account the structural complexity of transportation networks. Here we explore systematically the topological limitations of urban bicycle network development. For 62 cities we study different variations of growing a synthetic bicycle network between an arbitrary set of points routed on the urban street network. We find initially decreasing returns on investment until a critical threshold, posing fundamental consequences to sustainable urban planning: Cities must invest into bicycle networks with the right growth strategy, and persistently, to surpass a critical mass. We also find pronounced overlaps of synthetically grown networks in cities with well-developed existing bicycle networks, showing that our model reflects reality. Growing networks from scratch makes our approach a generally applicable starting point for sustainable urban bicycle network planning with minimal data requirements.
\end{abstract}

{\vskip 10mm}
  \end{@twocolumnfalse}
]

\footnotetext{*Corresponding author. Email: \href{mailto:misz@itu.dk}{misz@itu.dk}}

\section*{Introduction}

\begin{figure}[b!]
\centering
\includegraphics{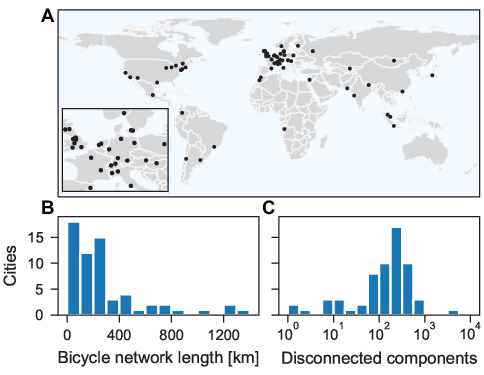}
\caption{\textbf{The state of existing bicycle networks.} (\textbf{A}) We extract street networks from 62 cities covering different regions and cultures; many are considered modern and well developed. (\textbf{B}) The distribution of city-wide lengths of bicycle tracks indicates negligible existing cycling infrastructure that is also (\textbf{C}) split into hundreds of disconnected components. See more details in \sitablename~1. Map created with: \url{https://github.com/mszell/bikenwgrowth} (v.1.0.0)}
\label{fig:cities}
\end{figure}

Cities worldwide are scrambling for sustainable solutions to their inefficient, car-centric transport systems \cite{banister2005unsustainable,nieuwenhuijsen2016cfc}. One promising, time-tested candidate is cycling. It is an efficient mode of sustainable urban transport that can account for the majority of intra-urban trips which are primarily short or medium-distance \cite{alessandretti2020shm}. Cost-benefit analysis that accounts for health, pollution, and climate, reveals that in the EU alone cycling brings a yearly benefit worth \euro\,24 billion while automobility costs society \euro\,500 billion \cite{gossling2019sca}. These insights provide further impetus for coordinated efforts to extend cycling infrastructure as one solution to the urban transport crisis and to effectively fight climate change \cite{gossling2020cnt,szell2018cqv}. Apart from being effective, this solution is also considerably more economic and wide-ranging than merely focusing on motor vehicle electrification \cite{creutzig2015trc,milovanoff2020elv,brand2021ccm}.

In practice, however, bicycle infrastructure development struggles with a political inertia due to the deep-rooted complexity of car-dependence \cite{mattioli2020political,feddes2020hard}: For example, Copenhagen took 100 years of political struggles to develop a functioning grid of protected on-street bicycle networks \cite{carstensen2015sdc} that continues to be split into 300 disconnected components today \cite{natera2020dso}. Accordingly, the most developed, influential bicycle network planning guidelines, such as the Dutch CROW manual \cite{crow2016dmb}, acknowledge that building up bicycle networks happens typically through decades-long, piecewise refinements. Unfortunately, there is overwhelming scientific consensus that the possible exit scenarios from the planetary climate crisis compatible with the $1.5^\circ$ goal are closing rapidly \cite{ripple2019world,ipcc2021ar6}. Given that transport is the most problematic sector \cite{lamb2021rtd} and that the majority of humanity is living in cities, making urban transport sustainable is therefore one of the most urgent societal issues \cite{creutzig2015trc,banister2005unsustainable,caiazzo2013air,mattioli2020political}. Electric cars are a potential solution to exhaust pollution but come with the same unavoidable downsides as traditional cars concerning urban livability \cite{nieuwenhuijsen2016cfc}, space allocation \cite{szell2018cqv}, road safety \cite{klanjcic2021iuf}, particulate matter pollution that is mainly caused by non-exhaust emissions \cite{jeong2022cpt}, public health and equity \cite{gossling2020cnt,pereira2017distributive}, among others. In particular, a sole focus on electric vehicles is counterproductive and ``active travel should be a cornerstone of sustainability strategies, policies and planning'' \cite{brand2021ccm}. Because of the fact that boosting active travel in cities has some of the highest potential to mitigate climate change and to improve public health \cite{gossling2020cnt,creutzig2015trc}, in this paper we focus on bicycle network development. While there has been historical political inertia in growing bicycle networks, the ongoing COVID-19 pandemic has prompted several cities to engage in successful accelerated network development, proving that such efforts are indeed possible \cite{lovelace2020mpp,kraus2021pci} (apart from already existing examples of fast growth \cite{marques2015hip,paris2021}). A systematic exploration of city-wide, comprehensive development strategies is therefore urgently needed.

Although the prevailing, piecewise application of qualitative policy guidelines in existing bicycle network planning \cite{zhao2018bip,crow2016dmb} might have a good track record in e.g.~Dutch cities and Copenhagen \cite{carstensen2015sdc}, this process lacks rigorous scrutiny: Are the resulting networks truly optimal? Can such policies be replicated in other cities? And are there fundamental topological limitations for developing a bicycle network? Indeed, an evidence-based, scientific theory of bicycle network development is missing.

There is a growing academic literature on analyzing bicycle networks of specific cities, for instance Montreal \cite{boisjoly2020bnp}, Seattle \cite{lowry2017qbn}, or recent data-driven approaches for Bogota \cite{olmos2020dcf}, London \cite{palominos2020ica}, or Berlin \cite{medeiros2021spatiotemporal}. While such studies are invaluable in terms of local enhancements and data consolidation for a particular place, here we instead focus on a global analysis, in particular on the \emph{fundamental topological limitations} of bicycle network development that are relevant for all urban environments, independent of the availability of traffic flow data \cite{mahfouz2021rsp}. This approach follows the idea of a \emph{Science of Cities} \cite{batty2013nsc} where we study the topological properties of bicycle networks that are \emph{independent of place} using computational, quantitative methods of \emph{Urban Data Science} \cite{resch2019hds}.

The vast majority of cities on the planet has negligible infrastructure for safe cycling \cite{szell2018cqv}. Indeed, urban transport infrastructure development worldwide has been heavily skewed towards automobiles since the 20\textsuperscript{th} century, today featuring well connected networks of streets for motorized vehicles \cite{natera2020dso}. Rather than uprooting the existing infrastructure and replacing it with an entirely new one---an economically infeasible strategy---we investigate how to retrofit existing streets into bicycle networks. Sacrificing specificity for generalizability, our formulation contains as a starting point two ingredients: the existing  street network of a city, and an arbitrary set of seed points. With these minimal ingredients we explore different growth strategies that sequentially convert streets that were designed for only cars to streets that are safe for cycling \cite{crow2016dmb,teschke2012route}. Using the CROW manual as a key reference and inspiration \cite{crow2016dmb}, the objective of all explored strategies is to create \emph{cohesive} networks, i.e.~well connected networks that cover a large fraction of the city area (see Materials and Methods).

Across the realistic strategies we report a growth phase that initially leads to a diminishing trend of quality indicators, until a critical fraction of streets are converted, akin to a percolation transition observed in critical phenomena and also present in the growth of other forms of transportation infrastructure as well as patterns of traffic~\cite{erdos1959rg, Zeng_2019, Gross_2020, olmos2020dcf, rhoads2020pso, mahfouz2021rsp}. In other words, initial investments into building cycling-friendly infrastructure leads to diminishing returns on quality and efficiency until the emergence of a well-connected giant component. Once this threshold is reached, the quality improves dramatically, to an extent which depends on the specific growth strategy. We provide empirical evidence that the majority of cities effectively lie below the threshold which might be hindering further growth, implying fundamental consequences to sustainable urban planning policy: To be successful in developing well-connected bicycle networks, cities must invest with the right growth strategy, and \emph{persistently}, to surpass a critical mass.

\section*{Results}

\begin{figure}[t!]
\centering
\includegraphics{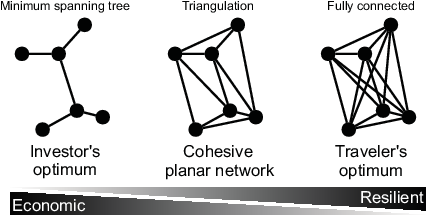}
\caption{\textbf{Optimal connected network solutions.} Adapted from Ref.~\cite{vannes2002dmt}. (Left) The investor's optimal strategy for a connected network is to invest as little as possible, minimizing total link length \cite{natera2020dso}. Its solution is a minimum spanning tree, maximally economic but minimally resilient with low directness, inadequate for travelers. (Right) The traveler's optimum connects all node pairs creating all direct routes. This solution is minimally economic, maximally resilient and direct, inadequate for investors. It also has crossing links and is therefore not a planar network. (Center) A both economic and resilient, as well as cohesive planar network solution in-between is the triangulation. In particular the minimum weight triangulation, approximated by the greedy triangulation, minimizes investment.}
\label{fig:mstfull}
\end{figure}

The starting point for our analysis is the manually sampled street networks of 62 cities aiming to capture a diversity of cultural regions and a large range of populations, population densities, areas, and network lengths, selected from cities where there is relatively complete data available \cite{barrington2017wur}, see Fig.~\ref{fig:cities}A and \sitablename~1. Here, links represent streets and nodes are street intersections. Being embedded in a metric space, these constitute planar graphs \cite{barthelemy2011sn}. We downloaded and processed these networks from OpenStreetMap using OSMnx \cite{boeing2017osmnx} (see Materials and Methods). 

Although many of the covered cities are from well developed regions, we observe that they have negligible bicycle infrastructure, Fig.~\ref{fig:cities}B. Additionally these are split into hundreds of disconnected components, Fig.~\ref{fig:cities}C, which has previously prompted analysis of strategies to merge them \cite{natera2020dso}. Although such strategies make sense in cities with already well established bicycle infrastructure, they are less useful in most other cities. Further, they produce minimum spanning tree-like solutions that are economically attractive but lack resilience and cohesion (Fig.~\ref{fig:mstfull}), and they potentially reinforce socioeconomic inequalities by connecting only already developed areas while ignoring under-developed ones \cite{mahfouz2021rsp}, prompting us here to grow new networks from scratch instead. By resilience we mean a general level of fault-tolerance \cite{zhang2015arn}: A resilient bicycle network should provide an acceptable level of service in the face of faults and challenges to normal operation, for example interruptions due to road works. The removal of a small fraction of links should not have a substantial impact on network metrics.

\begin{figure*}[p!] 
\vspace{-2.25cm}
\centering
\includegraphics{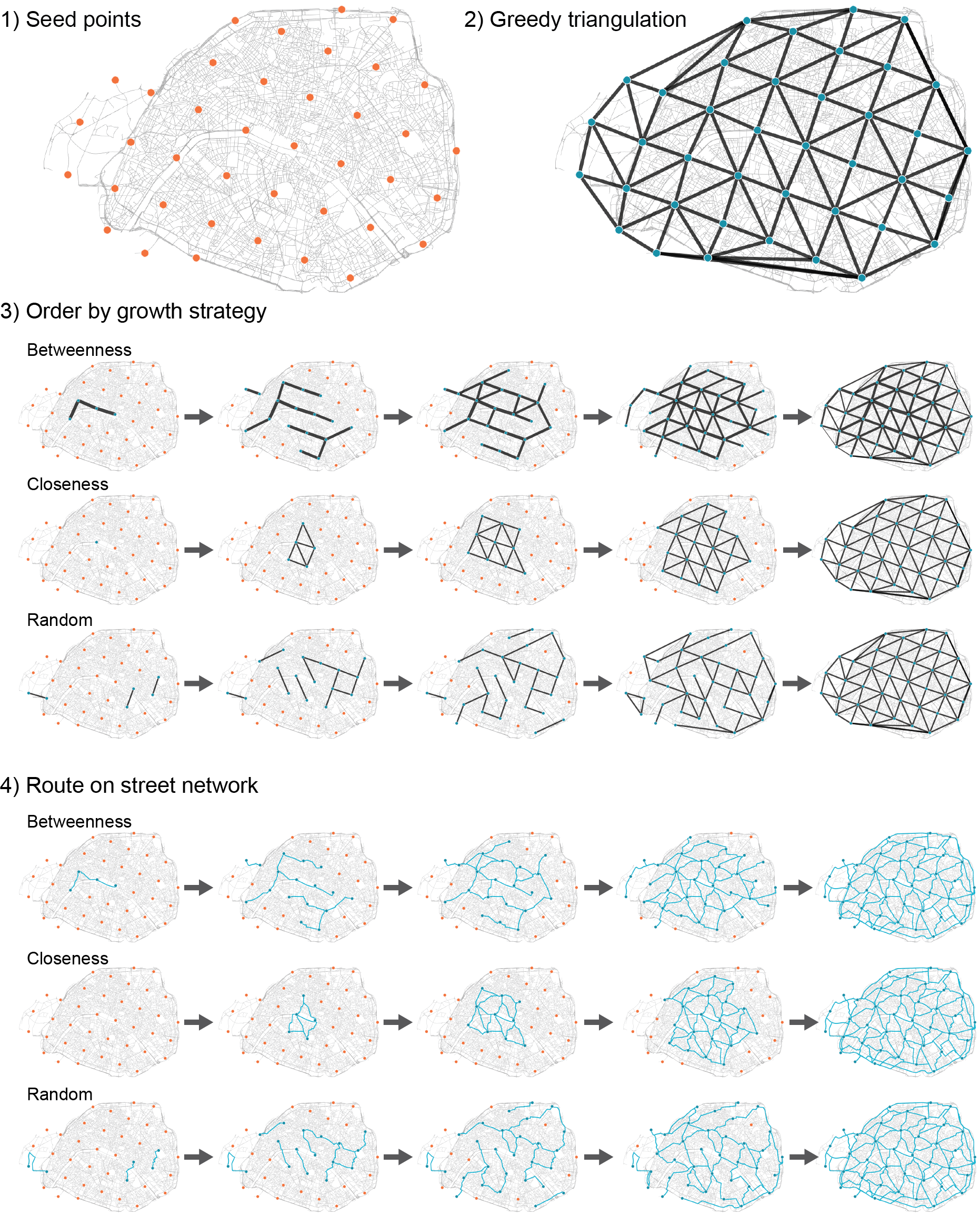}
\caption{\textbf{Growing bicycle networks.} Explorable interactively at: {https://growbike.net}. Illustrated here for Paris. Step 1) Seed points: A set of seed points (orange dots) is snapped to the intersections of the street network. Shown are grid points, alternatively we investigated rail stations. Step 2) Greedy triangulation: The seeds are ordered by route distance and connected stepwise without link crossings. Reached seeds are colored blue. Step 3) Order by growth strategy: One of three growth strategies (betweenness, closeness, random) is used to order the triangulation links from the strategy's 0-quantile (empty graph) to its 1-quantile (full triangulation), resulting in 40 growth stages. Shown are the five quantiles $q = 0.025, 0.125, 0.25, 0.5, 1$. Step 4) Route on street network: The links in the growth stages are routed on the street network. These synthetic bicycle networks are then analyzed for all 62 cities. Maps created with: \url{https://github.com/mszell/bikenwgrowth} (v.1.0.0)}
\label{fig:method}
\end{figure*}

\subsection*{Growing bicycle networks from scratch}
Our process of growing synthetic bicycle networks consists of four steps, Fig.~\ref{fig:method}, starting with the street network and the seed points. For an intuitive, interactive exploration see \mbox{https://growbike.net}.

\paragraph*{Step 1) Seed points.} 
An arbitrary set of seed points is snapped to the intersections of the street network. We investigate two versions of seed points: i) arranged on a grid, and ii) rail stations. Generally these seeds could have arbitrary coordinates, but in the CROW manual's context of origin-destination links \cite{crow2016dmb} they could represent points of interests such as district centers, shopping areas, schools, etc.

\paragraph*{Step 2) Greedy triangulation.}
All pairs of seed nodes are ordered by route distance and connected stepwise as the crow flies. A link is added only if it does not cross an existing link. This greedy triangulation is an easily computable proxy for the NP-hard minimum weight triangulation \cite{mulzer2008mtn}. It creates an approximatively shortest and locally dense planar network \cite{cardillo2006spp}, and a connected, cohesive, and resilient network solution minimizing investment, therefore satifsying both traveler and investor demands, Fig.~\ref{fig:mstfull}.

\paragraph*{Step 3) Order by growth strategy.}
Each of three growth strategies is used to order the greedy triangulation links from the strategy's 0-quantile (empty graph) to its 1-quantile (full triangulation), resulting in a sequence of growth stages. To study this growth process in a high enough resolution we split the growth quantiles into 40 parts $q = 0.025, 0.05, \ldots, 0.975, 1$. The three strategies are:
\begin{enumerate}
    \item \textbf{Betweenness} -- orders by the number of shortest paths that go through a link. It can be interpreted as the simplest proxy for traffic flow (assuming uniform traffic demand between all pairs of nodes). Thus, growing by betweenness is an approach that aims to prioritize flow. 
    \item \textbf{Closeness} -- starts with the ``most central'' node, i.e.~the node that is closest to all other nodes. From this seed, the network is built up by connecting the most central adjacent nodes. This approach is the most local approach possible and leads to a linear expansion of a dense as possible network from the topological city center. 
    \item \textbf{Random} -- adds links randomly and is used as a baseline. This strategy is not just a theoretical null model but well resembling how cities build their bicycle networks in practice, as we discuss later.
\end{enumerate}

\paragraph*{Step 4) Route on street network.} 
The abstract links in the 40 stages are made concrete: They are routed on the street network. These synthetic bicycle networks are then analyzed for all 62 cities.

\subsection*{Different growth strategies optimize different quality metrics}
We measure several network metrics to assess the quality of the synthetically grown networks and to compare them with existing bicycle networks. These metrics are: length $L$, length $L_\mathrm{LCC}$ of the largest connected component (LCC), coverage $C$, seed point coverage $C_\mathrm{seed}$, directness $D$, number of connected components $\Gamma$, global efficiency $E_\mathrm{glob}$, local efficiency $E_\mathrm{loc}$. We define coverage as the area of all grown structures endowed with a $500\,$m-buffer, see light blue areas in Fig~\ref{fig:grownvsexisting}B for an illustration. Directness measures the average ratios of euclidian distances versus shortest path distances on the network, while global efficiency provides a similar measure that accounts for disconnected components \cite{latora2001ebs}. See Materials and Methods for technical details.

\begin{figure*}[t!]
\centering
\includegraphics{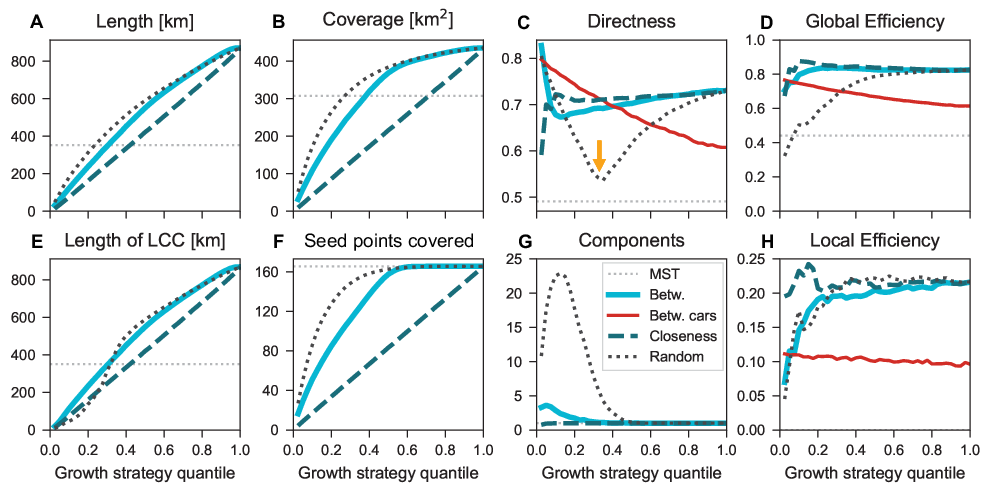}
\caption{\textbf{Different growth strategies optimize different network quality metrics.} The three thick curves show the changes of network metrics with growth following three strategies (betweenness, closeness, or random) averaged over all 62 cities for grid seeds. By construction all curves arrive at the same endpoint, but they develop distinctly before that. For rail seeds and individual cities see \sifigname{s}~5-10. Red curves show the car network's simultaneous decrease of quality metrics if a five times decrease of speed limits is assumed for cars along all affected streets. Grey dotted lines show metrics for the minimum spanning tree (MST) that connects all seeds with minimal investment. Growth of (\textbf{A}) length, (\textbf{B}) coverage, (\textbf{C}) directness, (\textbf{D}) global efficiency, (\textbf{E}) length of LCC, (\textbf{F}) seed points covered, (\textbf{G}) connected components, (\textbf{H}) local efficiency. The yellow arrow highlights the substantial dip in directness until the critical threshold which is more pronounced for random growth than for betweenness growth.}
\label{fig:growthavgcitygrid}
\end{figure*}

We first investigate how these quality metrics change throughout the growth process averaged over all cities, Fig.~\ref{fig:growthavgcitygrid}. The three thick curves show the change of the metrics with growth following the three strategies (betweenness, closeness, or random) for grid seeds. Similar results hold for rail stations, see \sifigname{s}~5-7. By construction all curves reach exactly the same point at the 1-quantile (full triangulation), but their development differs substantially before that. The minimum spanning tree (MST) solution is depicted as a baseline (grey dotted lines). This is the most economic connected solution that reaches all seeds, see Fig.~\ref{fig:mstfull}; therefore any connected solution that reaches all seeds must be at least as long as the MST.

From Fig.~\ref{fig:growthavgcitygrid}A we observe that length grows linearly for closeness and slightly faster for the other strategies because closeness prioritizes close links which typically have similar length, while betweenness and random growth selects distant links earlier. Random growth adds single links scattered randomly across the city and therefore has the fastest growth of coverage, Fig.~\ref{fig:growthavgcitygrid}B, followed by the betweenness-based strategy, while closeness leads to a linear growth. Directness, Fig.~\ref{fig:growthavgcitygrid}C, displays a large dip for random growth, from $D\approx 0.8$ down to $D\approx 0.53$ at the 0.345-quantile, and a smaller dip from $D\approx 0.83$ to $D\approx 0.68$ for betweenness growth at the 0.1-quantile. Directness starts lower for closeness growth, around $D \approx 0.59$ but quickly overtakes the other strategies at the 0.05-quantile. Global efficiency, Fig.~\ref{fig:growthavgcitygrid}D, starts at a high level, around $E_\mathrm{glob}\approx 0.7$, and grows slightly until $E_\mathrm{glob} \approx 0.82$ for both betweenness and closeness. Random growth starts instead much lower, around $E_\mathrm{glob} \approx 0.33$. The length of the LCC, Fig.~\ref{fig:growthavgcitygrid}E, is almost identical as the growth of length for betweenness and closeness because the LCC makes up most of the network here. However, the LCC in random growth has a sigmoid growth pattern as it takes longer for the components to connect, Fig.~\ref{fig:growthavgcitygrid}G. Coverage of seeds, Fig.~\ref{fig:growthavgcitygrid}F, is similar to coverage but more pronounced for random and betweenness growth. On average all seeds are covered before the 0.6-quantile. Finally, local efficiency, Fig.~\ref{fig:growthavgcitygrid}H, is steady around $E_\mathrm{loc} \approx 0.22$ for closeness, but grows fast for both betweenness and random growth from around $E_\mathrm{loc} \approx 0.05$.

To summarize, the different growth strategies optimize different quality metrics and come with different tradeoffs: 1) Use betweenness growth for fast coverage, intermediate connectedness and directness, and low local efficiency. 2) Use closeness growth for optimal connectedness and local efficiency but slow coverage. 3) Use random growth for fastest coverage but low directness, connectedness, and efficiency.

\begin{figure}[t!]
\centering
\includegraphics{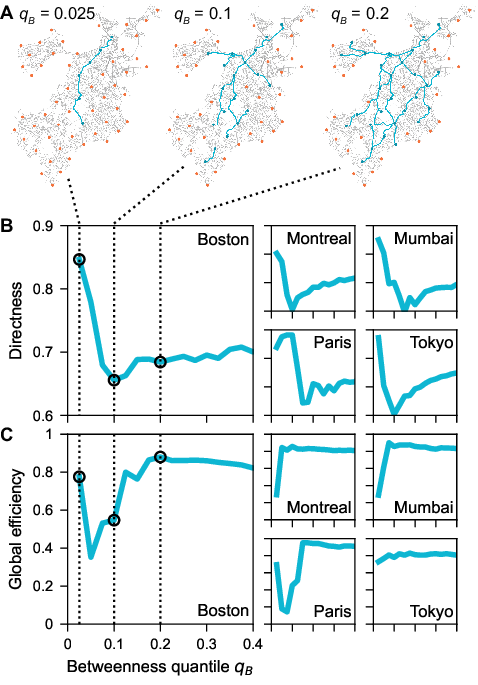}
\caption{\textbf{Network consolidation: Bicycle network growth has a dip of decreasing directness.} (\textbf{A}) Three early stages of betweenness growth in Boston. (\textbf{B}) Directness  sharply decreases initially due to tree-like growth (compare $q_{B}=0.025$ and $q_{B}=0.1$ for Boston). Once directness has reached a minimum ($q_{B}=0.1$), it starts growing slowly due to the appearance of cycles ($q_{B}=0.2$). The process is similar for the other cities (shown here for Montreal, Mumbai, Paris, Tokyo) and also holds for random growth, see \sifigname{s}~5,7,8,10. (\textbf{C}) We find mixed results for global efficiency: Mumbai and Montreal display a single jump, Tokyo is flat, while Boston and Paris shown an initial dip before increasing. Maps created with: \url{https://github.com/mszell/bikenwgrowth} (v.1.0.0)}
\label{fig:percolation}
\end{figure}

\subsection*{Network consolidation and non-monotonic gains in quality}
The dips observed in directness, see yellow arrow in Fig.~\ref{fig:growthavgcitygrid}C for random growth, are akin to a phase transition in a percolation process from a disconnected set of components to a sudden emergence of a giant connected component, as known for e.g.~random Erd\H{o}s-R\'enyi networks \cite{erdos1959rg}. Similar transitions have been observed in generalized network growth~\cite{achlioptas2009epr,bollobas1987tf}, including in various random spatial networks \cite{barthelemy2011sn} and sidewalk networks~\cite{rhoads2020pso}, and similar flavors of bicycle network growth~\cite{olmos2020dcf}. Figure~\ref{fig:percolation}A and B illustrates this consolidation process for individual cities: Links are added one by one, growing the largest connected component until a critical threshold at the curve's minimum (at $q_B=0.1$ for Boston), at which the largest connected component consolidates the majority of the network and starts forming cycles that in turn increase directness. Because connectedness increases around the critical threshold, the evolution of connected components is inverse to the evolution of directness, Fig.~\ref{fig:growthavgcitygrid}G. While the global efficiency averaged over all cities shows an initial increase followed by saturation, see Fig.~\ref{fig:growthavgcitygrid}D, we find mixed trends at the level of individual cities: Mumbai and Montreal track the average trend, Tokyo has a flat global efficiency, while Boston and Paris show a dip before the critical threshold is reached with rapid gains thereafter (Fig.~\ref{fig:percolation}C).

This network consolidation has important implications for policy and planning. The point at which the transition happens represents substantial investments into building the network. Stopping investments and growth \emph{before} this point leads to a net loss in investment as measured by infrastructure quality. Indeed, pushing past this threshold leads to substantive gains.

\subsection*{The effect of bicycle network growth on the street network}
While it is beneficial for a city to grow its bicycle network, it is important to ask how this growth affects the network of streets used by cars. The magnitude of this effect depends not only on the network topology, but also on the concrete bicycle infrastructure being implemented: shared spaces, unprotected cycle lanes, protected cycle tracks, bicycle streets, their width, and so on. To consider these factors, leading bicycle planning manuals consider a plethora of local variables \cite{crow2016dmb,nacto2014ubd}, for example road category, speed limit, volume of the motorized traffic, or car parking facilities. Therefore, to be conservative in our estimations, here we consider the strongest possible effect of new bicycle infrastructure on streets apart from complete replacement: We assume that all infrastructure would be built, for example, as a child-friendly ``fietsstraat'' or living street, i.e.~as a shared traffic space where cyclists and pedestrians have priority and cars are tolerated to pass through in walking speed \cite{crow2016dmb}. This assumption roughly translates to a reduction of speed limits for cars along the affected road sections by a factor of 5, for example from $50\,\mathrm{km/h}$ to $10\,\mathrm{km/h}$ or from $30\,\mathrm{km/h}$ to $6\,\mathrm{km/h}$. In our technical calculations we implemented a computational equivalent to this speed reduction -- an increase of the affected road section lengths by a factor of 5 \cite{crow2016dmb}. So, for example, for calculating directness along an affected road section, a street segment of length $100\,\mathrm{m}$ would then count as being $500\,\mathrm{m}$ long.

\begin{figure}[t!]
\centering
\includegraphics{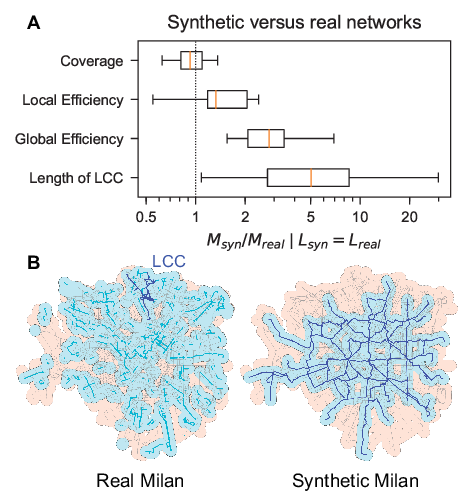}
\caption{\textbf{Synthetic bicycle networks perform several times better than existing ones.} (\textbf{A}) We plot the distributions (over cities) of the ratios $M_{\mathrm{syn}}/M_{\mathrm{real}}$ between network metrics of synthetic and existing topologies fixed at same length ($L_{\mathrm{syn}} = L_{\mathrm{real}}$), for betweenness growth and grid seeds (for all other growths see \sifigname~2). Synthetic networks have on average 5 times larger LCCs, 3 times the global efficiency, and higher local efficiency. Existing networks only tend to have better coverage because they are more scattered. (\textbf{B}) Illustration of high coverage (light blue area) due to extreme scattering and low length of LCC (dark blue sub-network) for Milan's existing bicycle network, versus its synthetic version at same length ($185\,\mathrm{km}$ at $q_B = 0.425$). The LCC for synthetic Milan is the whole network. Maps created with: \url{https://github.com/mszell/bikenwgrowth} (v.1.0.0)}
\label{fig:grownvsexisting}
\end{figure}

Given this strong constraint, we find the metric that is most affected is directness. It decreases approximately linearly with the bicycle network's growth, from $D \approx 0.8$ to $D \approx 0.6$, as the network is grown using the betweenness strategy (red curve in Fig.~\ref{fig:growthavgcitygrid}C). In other words, in the absence of any bicycle infrastructure, car-routes deviate by around 25\% from the Euclidean distance between any origin-destination, whereas once the full bicycle infrastructure is established, this increases to around 66\%. At around the 0.4 betweenness quantile, the directness of the bicycle network exceeds that of the car network. The global efficiency decreases from around $E_\mathrm{glob} \approx 0.75$ to $E_\mathrm{glob} \approx 0.6$, (red curve in Fig.~\ref{fig:growthavgcitygrid}D), while the local efficiency decreases negligibly from $E_\mathrm{loc} \approx 0.11$ to $E_\mathrm{loc} \approx0.10$ (red curve in Fig.~\ref{fig:growthavgcitygrid}H). Growing the bicycle network has no effect on the length and coverage of the automobile network, given that cars can still access all points on the street network, albeit in a longer time than they would without the bicycle infrastructure. We find almost identical behavior for the closeness and random growth strategies (\sifigname4). 

One of the effects of modifying the street infrastructure is the redistribution of load on the street intersections, measured by the betweenness centrality. It has been shown that while the global distribution of the betweenness centrality remains unchanged due to change in density of streets, the spatial distribution and clustering of the high betweenness nodes tend to change, thus redistributing areas of higher traffic~\cite{Kirkley_2018}. Two measures to quantify this effect are the spatial clustering and the anisotropy of the high betweenness nodes (see Materials and Methods). We find a slight increase (around 5\%) in spatial clustering and anisotropy for nodes in the 90th percentile of betweenness values but the effect is marginal (\sifigname~4).

\subsection*{Comparing synthetic with existing network metrics}
Although the growth processes described here are somewhat artificial, given the lack of accounting for practical limitations of bicycle network design---street width, incline, or political feasibility for instance---it is nevertheless prudent to compare the synthetic network with existing bicycle networks to gauge their general correspondence. To have a fair comparison in terms of length (which is a proxy for cost), we first select all cities that have a protected bicycle network with shorter length $L_{\mathrm{real}}$ than the fully grown synthetic network $L_{\mathrm{syn}}$ (42 out of 62 cities), and for each of them we fix the growth quantile where the synthetic length is equal to the real length, $L_{\mathrm{syn}} = L_{\mathrm{real}}$. Given this set of bicycle network pairs -- real versus synthetic at same length -- we then measure the ratio $M_{\mathrm{syn}}/M_{\mathrm{real}}$ between the synthetic quality metric $M_{\mathrm{syn}}$ and the quality metric of the existing infrastructure $M_{\mathrm{real}}$. The results for the metrics of coverage, local efficiency, global efficiency, and length of LCC are reported in Fig.~\ref{fig:grownvsexisting}A.

We find that synthetic networks have on average 5 times larger LCCs, 3 times the global efficiency, and higher local efficiency. Existing networks only tend to have better coverage because they are more scattered, as illustrated in Fig.~\ref{fig:grownvsexisting}B for Milan which has 230 disconnected components. Milan's scattered network provides an important lesson: Mere measures of total length or coverage are misleading when it comes to an efficient and safe infrastructure if the network is not well connected. Instead, if city planners were to develop and implement bicycle networks holistically, considering a city-wide rather than piece-wise local approach, much higher quality infrastructure could be derived to the benefit of the residents. Many examples such as Dutch cities, Seville, or Paris have already proven that this is indeed a realistic approach \cite{marques2015hip,schepers2017dutch,paris2021}.

For completeness we also compare our results to the closeness and random growth approaches, \sifigname~2. For closeness we find almost the same result as for betweenness, only with notably worse coverage which is to be expected given how closeness grows the covered area as slowly as possible. For the random growth approach we find the same coverage as existing infrastructure, and around 2 times the global efficiency and length of LCC. At first blush, this implies that even a naive random growth strategy can perform better than existing ones. However, this could be due to a number of reasons: For instance, in the random growth process described here, segments are added over at least $1.7\,\mathrm{km}$ in each step, whereas in real cities, segments are added in a more scattered fashion and in varying lengths. Further, cities can have non-negligible off-street bicycle tracks, for example through parks, a feature not considered in our analysis.

\subsection*{Comparing synthetic with existing network overlaps}

\begin{figure}[t!]
\centering
\includegraphics{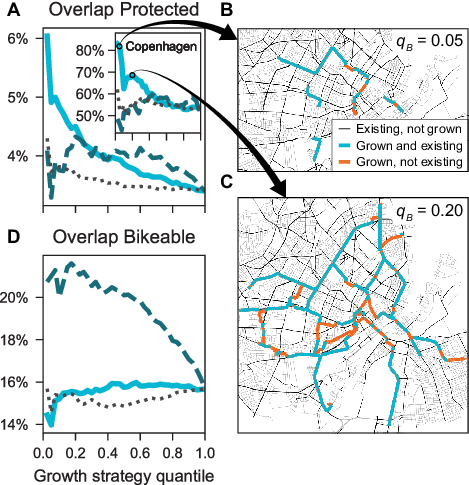}
\caption{\textbf{First stages of synthetic growth recreate existing networks.} Shown are results for rail station seeds averaged over all cities. Same legend as Fig.~\ref{fig:growthavgcitygrid}. (\textbf{A}) Growth by betweenness starts with high, then decreasing overlap with existing protected bicycle infrastructure. Inset: The effect is especially strong in cities with well developed on-street bicycle networks such as Copenhagen. Here the growth algorithm starts with over 80\% overlap. (\textbf{B}) Map of this high overlap in Copenhagen at the quantiles $q_{B}=0.05$ and (\textbf{C}) $q_{B}=0.20$. (\textbf{D}) The overlap with bikeable infrastructure has a notable effect only for growth by closeness due to traffic-calmed city centers: With increasing distance from the city center, overlap falls. Maps created with: \url{https://github.com/mszell/bikenwgrowth} (v.1.0.0)}
\label{fig:overlap}
\end{figure}

To gain a better understanding into how the synthetically grown parts compare to existing infrastructure, and thus the extent to which the growth models approximate reality, we measure the percent overlap of synthetic infrastructure with existing bicycle infrastructure. Figure~\ref{fig:overlap}A and B report for rail station seeds the average overlaps for protected cycle tracks and for bikeable infrastructure respectively, where bikeable infrastructure is defined as the union of protected tracks and streets with speed limits \mbox{$\leq 30\,\mathrm{km}/\mathrm{h}$} (see Materials and Methods). Results are qualitatively similar for grid seeds, see {\sifigname}s~8--10. 

For protected infrastructure, Fig.~\ref{fig:overlap}A, we find that growth by betweenness (solid line) starts with around $6\%$ overlap on average, then decreases fast reaching $3.5\%$. We find a similar behavior for random growth (dotted line) but with a smaller effect. We find no clear effect for growth by closeness (dashed line). These observations suggest that cities take into account flow (betweenness) when building their cycling infrastructure, and that rail stations play some role -- otherwise there would be no effect in the random growth. The betweenness overlap effect is especially strong in Copenhagen, Fig.~\ref{fig:overlap}A inset, which has a well-developed, cohesive on-street bicycle network. Figure~\ref{fig:overlap}B shows the synthetic growth stage at the second step, $q_{B}=0.05$. Remarkably, at this step over 80\% of the links suggested by the synthetic network model do already exist in reality. Even at $q_{B}=0.20$ there is almost 70\% overlap, Fig.~\ref{fig:overlap}C. These values are far higher than expected by chance: Given the length of Copenhagen's on-street cycle tracks we would expect only at most $24\%$ overlap in a random link placement.

The overlap for bikeable infrastructure looks different, Fig.~\ref{fig:overlap}D. Here, there is no clear effect for betweenness and random growth, but a very clear effect for closeness, which starts with high values (above $20\%$ on average), falling slowly to $16\%$. This observation is consistent with cities preferentially installing low speed limit areas in their city centers.

\section*{Discussion}
We grew synthetic bicycle networks in 62 cities following three different growth strategies all aiming to generate a cohesive network, i.e.~a network that is well connected and covers a large fraction of the city area. Studying the resulting networks we found a consistent critical threshold affecting directness in all cities and global efficiency in some of them, for the two most realistic strategies of growth by betweenness and random growth. This sudden network consolidation therefore has a fundamental policy implication: To grow bicycle networks successfully, cities must invest into them \emph{persistently}, to surpass short-term deficiencies until a critical mass of bicycle infrastructure has been built up. Further, from a topological perspective, cities should avoid traditional ``random growth-like'' strategies that follow local, stepwise refinement. Such strategies substantially shift the critical threshold, thereby hold up the development of a functional cycling infrastructure which could fuel adversarial objections to bicycle network expansions along the lines of ``We already built many bike tracks but nobody is using them, so why build more?" As we have shown, \emph{it is not a network's length that matters but how you grow it}.

Our main result focuses on directness because it is the most important metric for bicycle network planning apart from connectivity: It is the key metric to quantify network quality \cite{crow2016dmb}, and it is the best predictor or quantifiable policy aspect for adoption of cycling \cite{rietveld2004dbu,schoner2014mlb}. To ensure that our results are robust to other possible definitions of directness, we compared our main result, see Fig.~\ref{fig:growthavgcitygrid}C, using four different definitions, see Materials and Methods and \sifigname~3. Numerical values vary only insignificantly, all results are qualitatively identical for each definition, thereby establishing robustness.

By comparing metrics and overlaps of synthetic with existing networks, we gained an insight into the realism of our models. Our observations suggest that growth processes of existing protected bicycle networks contain a strong random ingredient and a detectable consideration for flow (betweenness) and rail stations. The random ingredient can be explained by the traditionally slow-paced urban planning processes arising from political inertia \cite{zhao2018bip,feddes2020hard}. Unfortunately, the random strategy is also the slowest in terms of network consolidation: It needs at least three times the investments than the betweenness strategy to reach the critical threshold. The rail effect can be explained by transit-oriented development efforts, where bicycle facilities are planned close to transit lines \cite{palominos2020ica,ibraeva2020tdr}. The remarkably high overlap with Copenhagen's well developed network suggests that our models could also be adapted to identify ``missing links'' in existing bicycle networks \cite{folco2022dbn,vybornova2022adm}.

Although the emergence of a giant component in network growth could have been anticipated with network science expertise, our results are not trivial: 1) This crucial insight is missing in the bicycle network planning manuals that practitioners use \cite{crow2016dmb}, 2) the different pros and cons of growth strategies have not been studied nor quantified before, 3) the policy dimension shows that reports on lengths and functionalities of under-developed bicycle networks must be scrutinized in an evidence-based way and take into account network structure. Only by being global and minimalistic, deliberately ignoring second order effects, does our approach uncover fundamental topological limitations of bicycle network growth independent of place. At the same time our results must be treated as \emph{statistical solutions}. By no means do they suggest concrete recommendations for new bicycle facilities, as a vast array of local idiosyncrasies (second order effects) would need to be accounted for \cite{nacto2014ubd,crow2016dmb}, including: road category, speed limit, volume of motorized traffic, or aspects of comfort \cite{quercia2014sph}. Despite the importance of these aspects, a transport network's geometry is its most fundamental limitation \cite{walker2018pcp} which is the reason we explored it first. Although our approach here is not yet aiming to provide concrete urban design solutions, it could be useful for planning purposes for easily generating an initial vision of a cohesive bicycle network -- to be refined subsequently \cite{folco2022dbn}. By publishing all our code as open source we facilitate such future refinements. Our minimal requirements on data are a deliberate limitation we impose for our framework to be applicable to data-scarce environments and thus to a large part of the planet \cite{barrington2017wur}: no lane widths, inclines, traffic flows, etc. are needed to optimize network topology.

The studied alternative approach of starting from rail station seeds instead from grid seeds seems reasonable, however care has to be taken to not amplify existing biases that are well-documented in the transport planning profession \cite{bullard2004hrt,hoffmann2016blw,pereira2017distributive}: For example, planning bicycle infrastructure only along metro stations that were built following elitist or racist biases would reinforce them, neglecting under-served regions and their inhabitants even further. The strength of our seed point approach lies in its arbitrariness that can bypass such issues: Grid seeds implement equal coverage and could be a starting point, to be refined carefully with e.g.~population density, traffic demand models, or flow data \cite{mahfouz2021rsp,jafino2021equity,folco2022dbn}. The biggest limitation of our approach is the sole focus on retrofitting street networks for safe cycling. This approach has some issues because it only considers on-street but no off-street bicycle infrastructure. We discuss the technical details of this limitation, mostly relevant for concrete bicycle network planning in low urban density, in \sinotename~1, concluding that future research on bicycle network growth should consider off-street solutions wherever possible.

Finally we discuss the effect of growing bicycle networks on limiting street networks for car traffic. Our flow analysis detects no substantial change of choke points. To be fair, this analysis is static and does not account for possibly nonlinear dynamic congestion effects which could be studied in arbitrary detail and precision. However, the state of the art in sustainable travel planning and systems design is clear that such short-term dynamics predictions are overtrumped by long-term behavioral effects \cite{nelson1997ibc,lyons2016gtp,itf2021tth,oecd2021tsn,european2004rcs}: Induced demand posits that the development of a functional cycling infrastructure will generally drive a modal shift towards cycling -- for latest evidence see, e.g.~Refs.\cite{marques2015hip,kraus2021pci} -- while the reclamation of ineffectively used automobile space will naturally lead to disappearing traffic. Therefore, the OECD recommends to replace the outdated ``predict and provide'' planning paradigm with the vision-led ``decide and provide'' principle \cite{lyons2016gtp,itf2021tth}. Our research follows this principle by prioritizing planning for access and the latent demand for cycling \cite{lovelace2017pct,marshall2021ttt} through a cohesive network, rather than optimizing hard to forecast flow dynamics that are trumped by stronger equilibrium effects in the long term.

Concerning the change from directness $D\approx0.8$ to $D\approx0.6$, it is unclear whether to interpret it as substantial or inconsequential. Following considerations of long-term systems design as above, we deem it more important to discuss whether a small or a large change is \emph{desired}. There are arguments for both sides: From the perspective of car-dependent transport planning the change should be small to not disrupt the existing system too abruptly \cite{mattioli2020political,lamb2020discourses,oecd2021tsn}. From the perspective of sustainability, human-centric urban planning, and climate research, the change should be large to boost efficient bicycle transport, livable cities, and to fight climate change effectively. Indeed, the CROW manual states that directness should be higher for cyclists than for cars \cite{crow2016dmb}. On top of that it could be argued that our eurocentric Copenhagen-style model of building a relatively sparse sub-network for cyclists is not going far enough, or that it could be out of place in other socio-cultural or land-use contexts \cite{cervero2009influences, hughes2012evolution, bijker1997bicycles}. For example, it could be inverted into a Barcelona-style model where dense patches of living streets -- Superblocks -- are built within a sparse sub-network of automobile arterials \cite{nieuwenhuijsen2020utp,oecd2021tsn}. In any case, resistance to such ideas needs to be anticipated \cite{gossling2020cnt,lamb2020discourses}, requiring vigorous policy making and a well-informed civil society following leading examples such as the Netherlands \cite{schepers2017dutch,feddes2020hard}. Sustainability science provides overwhelming evidence for the societal benefits of following such persistent implementations, facilitating the transition to cities with sustainable transport systems to counteract climate change effectively, and providing extraordinary benefits to public health and urban livability \cite{european2004rcs,nieuwenhuijsen2016cfc,klanjcic2021iuf,prietocuriel2021pte,caiazzo2013air,brand2021ccm}.

Summarizing limitations and future work, we call for network development models that combine both the long-term goal of a cohesive, accessibility-focused network as we do here, and the use of empirical, place-specific or street-level data for refinements \cite{folco2022dbn}, while being critical of flow-optimizing engineering approaches \cite{oecd2021tsn}. On a policy level, more research is needed into understanding socio-technical processes to overcome political inertia \cite{mattioli2020political,hughes2012evolution,bijker1997bicycles}. Finally, let us answer the questions posed in the beginning. Are bicycle networks of existing cities optimal? -- Our example of Milan has shown that in general, they are not, or that they are built in a too disconnected way. However, when it comes to well developed cities like Copenhagen, we find -- despite many still outstanding gaps \cite{vybornova2022adm} -- higher than expected overlap in the first growth stages, showing signs of an optimization process. Can optimal growth policies be replicated in other cities? -- Yes, the technical solutions exist, and the scale of investment is mostly a matter of political will as we can see from the Netherlands, Sevilla, or Paris \cite{marques2015hip,schepers2017dutch,paris2021}. And are there fundamental topological limitations for developing a bicycle network? Yes, there is a critical threshold to overcome until a functional bicycle network emerges. Because of this threshold and its dependence on the growth strategy, our practical recommendations are to concentrate investments as early as possible, and to grow for the whole city instead of piece-wise.

\section*{Materials and Methods}
\subsection*{Network data and growth}
\paragraph{Infrastructure networks.} We downloaded existing street and bicycle networks for 62 cities from OpenStreetMap (OSM) on 2021-02-26 using OSMnx \cite{boeing2017osmnx}. For each city, three networks were downloaded: Street network, protected bicycle network, bikeable network. Each node is an intersection, each link is a connection between two intersections. A protected bicycle network is the union of all OSM data structures that encode protected bicycle infrastructure, both on-street and off-street. Following the cycling safety literature, we consider only protected bicycle networks in our main analysis because safe cycling in general conditions is only ensured through physical separation from vehicular traffic  \cite{teschke2012route,crow2016dmb,szell2018cqv,nacto2014ubd,schepers2017dutch} We also consider for additional analysis the ``bikeable'' network, which is the union of a protected bicycle network and all streets with speed limits $\leq 30\,\mathrm{km}/\mathrm{h}$ or $\leq 20\,\mathrm{m}/\mathrm{h}$ (including living streets). In special conditions such street segments can be considered safe for cycling, but not in general \cite{crow2016dmb,teschke2012route}; safety is a complex topic requiring a deep discussion of a multitude of variables \cite{klanjcic2021iuf}, therefore we consider it outside the scope of this work. OSM data has been generally found to be of high quality and completeness \cite{haklay2010openstreetmap,barrington2017wur}, but multicity studies using bicycle infrastructure data such as ours could potentially suffer from some labeling inconsistencies, especially for less common types of bicycle infrastructure \cite{ferster2019openstreetmap}. 

\paragraph{Seed points.} Rail station seeds consist of all railway and metro stations. A few of the considered cities do not have rail stations. For creating grid seeds, we created grid points at a distance of $1707\,\mathrm{m}$, ensuring a tolerable average distance of $167\,\mathrm{m}$ (2 minutes walking) over the whole city to the triangulated network in the worst case, see \sinotename~2. We then rotated this grid to align it with the city's most common street bearing \cite{boeing2019uso}, and snapped the grid points to the closest street network intersections within a $500\,\mathrm{m}$ tolerance. The rotation is mostly important for US cities that have a grid-like street network, e.g.~Manhattan, for creating straight routes.

\paragraph*{Greedy triangulation.}
The greedy triangulation orders all pairs of nodes by route distance and connects them stepwise as the crow flies. A link is added only if it does not cross an existing link. This triangulation is a $O(N\log N)$ computable proxy for the NP-hard minimum weight triangulation \cite{mulzer2008mtn} with an approximation ratio of $\Theta(\sqrt N)$ \cite{levcopoulos1998qta}. The greedy triangulation is fast and solvable for any set of nodes. Computing a quadrangular grid, as suggested by the CROW manual \cite{crow2016dmb}, or a quadrangulation, is in general not possible for arbitrary sets of nodes and also computationally less feasible \cite{toussaint1995qps}.

\paragraph{Growth strategy: Betweenness centrality.} This is a path-based measure that computes the fraction of paths passing through a given node $i$ \cite{Freeman1977},
\begin{equation}
    C_{B}(i)=\frac{1}{N} \sum_{s \neq t} \frac{\sigma_{s t}(i)}{\sigma_{s t}}
\end{equation}
where $\sigma_{st}$ is the number of shortest paths going from nodes $s$ to $t$ and $\sigma_{st} (i)$ is the number of these paths that go through $i$.

\paragraph{Growth strategy: Closeness centrality.}  Measures the total length of the shortest paths from a node $i $ to all other nodes in the network \cite{Freeman1978},
\begin{equation}
    C_{C}(i)=\frac{N - 1}{\sum\limits_{i\neq j} d(i, j)}
\end{equation}

\subsection*{Network metrics}
\paragraph{Cohesion.}
The CROW manual \cite{crow2016dmb} describes qualitatively what it means for a network to be \emph{cohesive}: a ``combination of grid size and interconnection". It states that this is the most elementary requirement for a bicycle network but without a rigorous definition. We interpret this concept as having both high connectedness (few disconnected components) and coverage, see below. A cohesive network should also be resilient, see below, which excludes pathological cases like the minimum spanning tree.

\paragraph{Coverage.}
We measure spatial coverage of the network as the union of the $\varepsilon$-neighborhoods of all network elements, i.e.~a buffer of $\varepsilon\,\mathrm{m}$ around all links and nodes. Here we set $\varepsilon = 500\,\mathrm{m}$ together with the grid seed distance, as this implies a theoretical coverage of $100\%$ of the city area for a grid triangulation and an average distance to the network of $167\,\mathrm{m}$, see \sinotename~2. In general, a cohesive bicycle network should cover the majority of the city area.

\paragraph{Seed point coverage.}
This metric refers to the number of seed points that have been covered by network elements (by the coverage defined above).

\paragraph{Components.}
The number of disconnected components is the number of maximal connected subgraphs, i.e.~all pairs of nodes within one component are reachable with a path but there is no path between nodes from different components.

\paragraph{Directness.} The directness between two nodes $i$ and $j$ is generally defined as the ratio $\frac{d_{E}(i, j)}{d_{G}(i, j)}$ between euclidean distance $d_{E}(i, j)$ and shortest path distance $d_{G}(i, j)$. The average of this ratio over all pairs of nodes is then the directness of the whole network: 
\begin{equation}
    D = \left\langle \frac{d_{E}(i, j)}{d_{G}(i, j)} \right\rangle_{i \neq j}
\end{equation}
Node pairs $i$ and $j$ are considered from within the same components because directness is a meaningless concept for nodes from different components. Other possible definitions for directness could be: 
\begin{itemize}
\item The previous definition but only applied to the LCC: $D = \left\langle \frac{d_{E}(i, j)}{d_{G}(i, j)} \right\rangle_{i \neq j \in LCC}$
\item The ratio of total euclidian distances and shortest path distances: $D= \frac{\sum_{i \neq j} d_{E}(i, j)}{\sum_{i \neq j}d_{G}(i, j)}$
\item The previous definition but only applied to the LCC: $D= \frac{\sum_{i \neq j \in LCC} d_{E}(i, j)}{\sum_{i \neq j \in LCC}d_{G}(i, j)}$
\end{itemize}
We calculated directness according to all these different definitions as a robustness check, see \sifigname~3. Numerical values vary only insignificantly, all results are qualitatively identical for each definition.

\paragraph{Local and Global Efficiency.}
A network's global efficiency is defined as \cite{latora2001ebs}:
\begin{equation}
    E_{\mathrm{glob}}= \frac{\sum\limits_{i \neq j } \frac{1}{d_{G} (i,j)}}{\sum\limits_{i \neq j } \frac{1}{d_{E} (i,j)}}
\end{equation}
A network's local efficiency is defined as the average of global efficiencies $E_{\mathrm{glob}} (i)$ over each node $i$ and its neighbors,
\begin{equation}
E_{\mathrm{loc}}= \frac{1}{N} \sum_{i=1}^{N} E_{\mathrm{glob}} (i)
\end{equation}
Local efficiency measures local fault tolerance and therefore operationalizes the concept of resilience on a local level.

\paragraph{Spatial clustering and Anisotropy.}
We first specify a threshold $\theta$ and identify the $N_{\theta}$ nodes with high betweenness above the $\theta$-th percentile. Then, we compute their spread about their center of mass $$x_{c m}=\frac{1}{N_{\theta}}\sum_{i=1}^{N_{\theta}} x_{i}$$ where $x_{i}$ specifies their coordinates,
normalizing for comparison across networks of different sizes via
\begin{equation}
C_{\theta}=\frac{1}{N_{\theta}\langle X\rangle} \sum_{i=1}^{N_{\theta}}\left\|x_{i}-x_{c m}\right\|,
\end{equation}
where $$\langle X\rangle=\frac{1}{N}  \sum_{i=1}^{N} \| x_{i}-x_{c m} \|$$ is the average distance of all nodes in the network to the center of mass of the high betweenness cluster.

Transition between the topological and spatial regimes is quantified by the increasingly isotropic layout of the high betweenness nodes with increasing edge-density. The anisotropy factor is defined by the ratio
\begin{equation}
  A_{\theta}=\frac{\lambda_{1}}{\lambda_{2}}
\end{equation}
where $\lambda_{1} \leq \lambda_{2}$ are the positive eigenvalues of the covariance matrix of the spatial position of the nodes with betweenness above the threshold $\theta$ \cite{Kirkley_2018}.

For the largest 15 cities we calculated these values only at the 0, 0.5, and 1 quantiles of the growth strategies due to computational feasibility. Therefore, \sifigname~4 reports average values over the 47 smallest cities.

\bibliography{main}
\bibliographystyle{ScienceAdvances}

\noindent \textbf{Acknowledgements:}
We thank Cecilia Laura Kolding Andersen and Morten Lynghede for developing and implementing the visualization platform. We are grateful to Anders Hartmann, Anastassia Vybornova, and Laura Alessandretti for helpful discussions. We thank the ITU High-Performance Computing cluster for computing resources and support. We gratefully acknowledge the open data that this article is based on, from \url{https://www.openstreetmap.org}, copyright OpenStreetMap contributors.\\
\noindent \textbf{Funding:} This work was supported by the Danish Ministry of Transport.\\
\noindent \textbf{Author Contributions:} M.S.~designed the study with input from R.S.~and G.G. M.S.~wrote the manuscript with input from all authors. M.S.~acquired and pre-processed the data with input from S.M.~and T.P., and performed the simulations. M.S.~directed the project, R.S.~and G.G.~helped supervise the project. M.S.~measured the results, aided by S.M. T.P.~performed the analysis on grid size and network coverage. All authors discussed the results.\\
\noindent \textbf{Competing Interests:} The authors declare that they have no competing financial interests.\\
\noindent \textbf{Data and materials availability:} All code used in the research is open-sourced, available at: \url{https://github.com/mszell/bikenwgrowth}. All data used and generated in the research are publicly available at Zenodo \cite{michaelszell2021}: \url{https://zenodo.org/record/5083049}. Interactively growing networks, plots and video visualizations for all 62 cities can be explored and downloaded at the accompanying visualization platform: \url{https://growbike.net}

\end{document}